\documentclass[10pt,a4paper]{article}
\usepackage{epsf}

\begin{document}

\vspace{0.5cm}

\begin{center}

{\bf \large    Neutrino Conversions in Active Galactic Nuclei\footnote{
 Talk given at 9th Lomonosov Conference on Elementary Particle Physics, 20-26 
 September, 1999, Moscow, Russia.}}\\
\vspace{0.4cm}
{Athar Husain\footnote{E-Mail: athar@phys.metro-u.ac.jp}}\\[0.4cm]
  Department of Physics, Tokyo Metropolitan University,\\
        Minami-Osawa 1-1, Hachioji-Shi, Tokyo 192-0397, Japan
        
\end{center}

\vspace{0cm}

\begin{abstract}

I discuss the possibility of production of high energy cosmic
neutrinos ($E\, \geq 10^{6}$ GeV) in cores of active galactic nuclei  and 
study some of the effects of 
neutrino mixing on their subsequent propagation. I also discuss the prospects
for observations of these high energy cosmic neutrinos in new km$^{2}$ 
surface area underwater/ice neutrino telescopes.

\end{abstract}

\section{Introduction}

	In this contribution  I discuss the possibility of production of 
high 
energy  cosmic  neutrinos ($E\, \geq 10^{6}$ GeV) in cores of Active Galactic
Nuclei (AGN) originating from proton acceleration, the effects of three 
 flavour neutrino mixing on these high 
energy cosmic neutrino fluxes and the prospects for their observations 
 in new km$^{2}$ surface area underwater/ice neutrino telescopes.

	In addition to AGNs, high energy cosmic neutrinos may also be 
produced in several currently envisaged other cosmologically distant
astrophysical sources. These sources may include, for instance, 
 Gamma Ray Burst fireballs and Topological Defects \cite{review}.
For some possible effects of neutrino 
mixing other than the flavour one on high energy cosmic neutrino fluxes, 
 see \cite{NPB}. The present study is particularly useful as several 
 high energy neutrino telescopes  are now at their rather advanced stage of 
development and deployment \cite{MOSCOSO}. 

	I start in Section 2 with a brief description of possibility of 
production of high energy cosmic neutrinos and discuss in some detail the 
 effects of three flavour neutrino
mixing on their subsequent propagation and further discuss the prospects for
their detection. In Section 3, I summarize the results.

\section{High Energy Neutrinos from AGNs}

\subsection{Production}

	High energy cosmic  neutrinos may mainly be produced either in 
$p\gamma $ or in $pp $ collisions in a cosmologically distant environment. 

In $p\gamma $ collisions,
high energy $\nu_{e}$ and $\nu_{\mu}$ are mainly produced through 
$p+\gamma \rightarrow \Delta^{+}\rightarrow n+\pi^{+}$ (typically with 
 $\nu_{e}/\nu_{\mu}\, \sim 1/2$). 
The same collisions will give rise to a greatly suppressed  
 high energy $\nu_{\tau}$  flux ($\nu_{\tau}/\nu_{e, \mu}\,<\, 10^{-5}$)
 mainly through 
 $p+\gamma \, \rightarrow \, D^{+}_{S}+\Lambda^{0}+\bar{D}^{0}$. 
 In $pp$ collisions, the $\nu_{\tau}$ flux may be obtained through 
$p+p\rightarrow D^{+}_{S}+X$.
The relatively small cross-section for $D^{+}_{S}$ production 
together with the low branching 
ratio into $\nu_{\tau}$ implies that the 
$\nu_{\tau}$ flux in $pp$ collisions is also suppressed up to 5 
orders of magnitude relative to $\nu_{e}$ and/or 
$\nu_{\mu}$ fluxes (which are mainly produced through $\pi^{\pm}$) 
\cite{ATHAR}. 

\subsection{Propagation}

	Matter effects on vacuum neutrino oscillations are
relevant if 
\linebreak 
 $G_{F}\rho /m_{N}\, \sim \, \delta m^{2}/2E$. Using $\rho$ 
from Ref. \cite{szabo} as an example, it turns out that matter effects are 
absent for $\delta m^{2}\, \geq \, {\cal O}(10^{-10})$ eV$^{2}$. 
Matter effects are not expected to be important in the neutrino 
production regions around AGN and will not be further discuss here.

	In the framework of three flavour analysis, the flavour precession
probability from $\alpha $ to $\beta $ neutrino flavour is \cite{book} 

\begin{eqnarray}
 P(\nu_{\alpha}\rightarrow \nu_{\beta})  \equiv  P_{\alpha \beta}  & = 
 & \sum^{3}_{i=1}|U_{\alpha i}|^{2}|U_{\beta i}|^{2}
                         \nonumber  \\
               &      & +\sum_{i \neq j} U_{\alpha i}U^{\ast}_{\beta i}
                        U^{\ast}_{\alpha j}U_{\beta j}
                        \cos\left(\frac{2\pi L}{l_{ij}}\right),
\end{eqnarray}
where $\alpha, \beta = e, \mu, $ or $\tau $. $U$ is the 3$\times $3 MNS mixing 
matrix and can be obtained in usual notation through 
\begin{eqnarray}
 U\, & \equiv & R_{23}(\theta_{1})
 \mbox{diag}(e^{-i\delta /2},1,e^{i\delta /2}) \nonumber \\ 
  & & \cdot R_{31}(\theta_{2}) \mbox{diag}(e^{i\delta /2},1,e^{-i\delta /2})
       R_{12}(\theta_{3}), 
\end{eqnarray}
thus coinciding with the standard form given by the Particle Data Group 
\cite{pdg}. In Eq. (1), $l_{ij}\simeq 4\pi E/
\delta m^{2}_{ij}$ with $\delta m^{2}_{ij} \equiv |m^{2}_{i}-m^{2}_{j}|$ and 
$L$ is the distance between the source and the detector. For simplicity, 
 I assume here a vanishing value for CP violating phase $\delta $ and 
 $\theta_{31}$ in $U$. 

	At present, the atmospheric muon and solar electron neutrino deficits 
 can be explained with 
oscillations among three active neutrinos \cite{RECENT}. For this, 
 typically, $\delta m^{2}
\sim {\cal O}(10^{-3})$ eV$^{2}$ and $\sin^{2}2\theta \sim {\cal O}(1)$ for 
the explanation of atmospheric muon neutrino deficit, 
whereas for the explanation of solar electron neutrino deficit, we may have  
$\delta m^{2} \sim {\cal O}(10^{-10})$ eV$^{2}$ and 
$\sin^{2}2\theta \sim {\cal O}(1)$ [just so] or $\delta m^{2}
\sim {\cal O}(10^{-5})$ eV$^{2}$ and $\sin^{2}2\theta \sim {\cal O}(10^{-2})$
[SMA (MSW)] or $\delta m^{2}
\sim {\cal O}(10^{-5})$ eV$^{2}$ and $\sin^{2}2\theta \sim {\cal O}(1)$ 
[LMA (MSW)]. The present status of data thus permits multiple oscillation 
 solutions to solar neutrino deficit. I intend to discuss here
implications of these mixings for high energy cosmic neutrino propagation.

	In the above explanations, the total range of $\delta m^{2}$ is 
$10^{-10}\leq \delta m^{2}/$ eV$^{2} \leq 10^{-3}$ irrespective of neutrino 
 flavour. The typical energy span relevant for possible flavour 
 identification for high
energy cosmic neutrinos is $2\cdot 10^{6}\leq E/$GeV$\leq 2\cdot 10^{7}$ in 
which currently the neutrino flux from cores of AGNs dominate. Taking a 
typical distance between the AGN and our galaxy as $L \sim 100$ Mpc (where 
1 pc $\sim 3\cdot 10^{16}$ m), note that $\cos $ term in Eq. (1) vanishes 
and so Eq. (1) reduces to 
 
\begin{equation}
   \langle P_{\alpha \beta} \rangle \simeq  
   \sum^{3}_{i=1}|U_{\alpha i}|^{2}|U_{\beta i}|^{2}.
\end{equation}

It is assumed here that no relatively dense objects exist between the AGN and 
the earth so as to effect significantly this oscillations pattern.
Note also that since $\langle P_{\alpha \beta }\rangle $ in above Eq. is 
 symmetric under the exchange of indices $\alpha $ and
$\beta $ implying that no $T$ (or $CP$) violation effects arise in 
neutrino vacuum flavour oscillations for high energy cosmic neutrinos 
\cite{cabibbo}.
 
	Let me denote by $F^{0}_{\alpha }$, the intrinsic neutrino fluxes. 
From the discussion in the previous Subsection, it follows that 
$F^{0}_{e} : F^{0}_{\mu} :F^{0}_{\tau} = 1 :2 : < 10^{-5}$. For simplicity, 
I take these ratios as 1 : 2 : 0. In order to estimate the final 
 (downward going) flux ratios of high energy cosmic neutrinos reaching on 
 earth, let me introduce a 3$\times $3 matrix of vacuum flavour 
precession probabilities such that 

\begin{equation}
 F_{\alpha} = \sum_{\beta}\langle P_{\alpha \beta}\rangle F^{0}_{\beta},
\end{equation}
where  the unitarity conditions for $\langle P_{\alpha \beta} \rangle $ read as

\begin{eqnarray}
 \langle P_{ee} \rangle +\langle P_{e\mu }\rangle +\langle P_{e\tau }
 \rangle &  = & 1,\nonumber \\
 \langle P_{e\mu} \rangle +\langle P_{\mu \mu }\rangle +\langle P_{\mu \tau }
 \rangle &  = & 1,\nonumber \\
 \langle P_{e\tau } \rangle +\langle P_{\mu \tau }\rangle +\langle 
 P_{\tau \tau }\rangle &  = & 1. 
\end{eqnarray}

The explicit form for the matrix $\langle P \rangle $ in case of 
just so flavour oscillations as solution to solar neutrino problem along with
the solution to atmospheric neutrino deficit in terms of $\nu_{\mu}$ to 
$\nu_{\tau}$ oscillations with maximal mixing is

\begin{equation}
 \langle P \rangle = \left( \begin{array}{ccc}
                             1/2 & 1/4 & 1/4 \\
                             1/4 & 3/8 & 3/8 \\
                             1/4 & 3/8 & 3/8 
                            \end{array}
                      \right).
\end{equation}
Using Eq. (6) and Eq. (4), it follows that $F_{e}: F_{\mu }: F_{\tau } = 1:
1: 1$ at the level of $F^{0}_{e}$. Also, Eq. (5) is satisfied. The same 
flux ratio is obtained in the remaining two cases for which the corresponding 
$\langle P \rangle$ matrics are: [for SMA (MSW)]

\begin{equation}
 \langle P \rangle = \left( \begin{array}{ccc}
                             1 & 0 & 0 \\
                             0 & 1/2 & 1/2 \\
                             0 & 1/2 & 1/2 
                            \end{array}
                      \right),
\end{equation}
whereas in case of LMA (MSW),

\begin{equation}
 \langle P \rangle = \left( \begin{array}{ccc}
                             5/8  & 3/16 & 3/16 \\
                             3/16 & 13/32 & 13/32 \\
                             3/16 & 13/32 & 13/32 
                            \end{array}
                      \right).
\end{equation}

 Thus, essentially independent of the 
oscillation solutions for solar neutrino problem, it follows that 
 $F_{e}: F_{\mu }: F_{\tau } = 1: 1: 1$. The deviations from these ratios are
estimated to be small \cite{polaska}.    	

	Summarizing, although intrinsically the downward going high energy 
 cosmic tau neutrino flux is negligibally small  
 however because of vacuum flavour 
oscillations it becomes comparable to $\nu_{e}$ flux thus providing some 
prospects for its possible detection.

\subsection{Prospects for detection}

	I briefly mention here the prospects for detection of downward going 
high energy cosmic tau neutrinos through double shower technique \cite{SANDIP}.
For prospects of  observations of high 
energy cosmic tau neutrinos other than double shower technique, see 
\cite{upward}, whereas for possibility of detection of non tau neutrinos, 
see \cite{GQRS}. 

The downward going tau neutrinos reaching close to the 
 surface of the detector may undergo  a charged current 
 deep inelastic 
scattering with nuclei inside/near the detector and produce 
 a tau lepton in addition
to a hadronic shower. 
\begin{figure}[t]
\leavevmode
\epsfxsize=3.5in
\epsfysize=2.4in 
\epsfbox{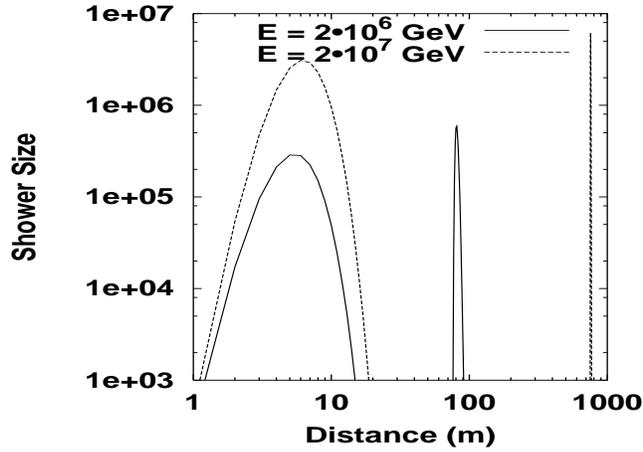}
\caption{Typical longitudinal development of a double shower  
         produced by the deep inelastic charged current interaction of the 
         tau neutrino in water/ice.}
\label{doubleshower}
\end{figure}
This tau lepton
traverses a distance, on average proportional to its energy, 
before it decays back into a tau 
neutrino and a second shower most often induced by decay hadrons. 
The second shower is expected to carry about twice as much energy 
as the first and such double shower signals are commonly referred 
to as a double bangs. 
As tau leptons are not expected to have further relevant 
interactions (with high energy loss) in their decay 
timescale, the two showers should be 
 separated by a clean $\mu$-like track.

The calculation of downward going contained but separable  
 double shower event rate can be carried out by  replacing 
 the muon range expression with the tau range expression and then 
 subtracting  it from the
linear size of a typical high energy neutrino telescope  
 in the event rate formula while using the expected 
$\nu_{\tau}$ flux spectrum given by Eq. (4). This ensures that the two 
 separate showers  are 
contained within km of  the underwater/ice  detector.
Here, I restrict myself by mentioning that the expected number of contained 
but separable double showers 
induced by downward going high energy tau neutrinos for
 $E\sim 2\cdot 10^{6}$ GeV
may be $ \sim {\cal O}(10) $/yr$\cdot$sr irrespective of the oscillation 
 solutions of solar 
neutrino problem, if one uses the $F^{0}_{e}$ from Ref. \cite{szabo} as an 
 example. At this energy,
the two showers initiated by the downward going high energy cosmic tau 
 neutrinos are well separated ($
\geq $ 70 m) such that the size of the second shower is essentially 2 
times
the first shower and the two showers are connected by a $\mu-$like track 
 (see Fig. 1). This 
identification, if empirically realized, may provide a possibility to isolate
$\nu_{\tau}$ flavour from the rest of neutrino flavors.
The chance of having double shower events induced by non tau 
neutrinos is negligibly small for relevant energies.  

\section{Conclusions}

	1. Intrinsically, the flux of high energy cosmic tau neutrinos is quite
small, relative to non tau flavour neutrinos, typically being 
 $F^{0}_{\tau}/F^{0}_{e, \mu}\, <\, 10^{-5}$ (whereas 
 $F^{0}_{e}/F^{0}_{\mu} \sim 1/2$) from cosmologically distant 
astrophysical sources, namely, for instance, cores of Active Galactic Nuclei.

	2. Because of neutrino oscillations, this ratio can be greatly 
 enhanced. In
the context of three flavour neutrino mixing scheme which can accommodate the 
oscillation solutions to solar and atmospheric neutrino deficits in terms of 
 oscillations between three active neutrinos, the final
ratio of fluxes of downward going high energy cosmic neutrinos on earth is 
 $F_{e}\sim F_{\mu}\sim F_{\tau} \sim
F^{0}_{e}$, essentially irrespective of the oscillation solutions to solar 
neutrino problem.

	3. This enhancement in high energy cosmic tau neutrino flux may lead 
to the possibility of its detection in km$^{2}$ surface area high energy 
neutrino telescopes. For $2\cdot 10^{6}\leq E$/GeV
$\leq 2\cdot 10^{7}$, the downward going high energy cosmic tau neutrinos
may produce a double shower signature because of charged current deep 
inelastic scattering followed by a subsequent hadronic decay of the associated
 tau lepton.

\section*{Acknowledgments}

I thank Japan Society for the Promotion of 
Science (JSPS) for financial support.

\end{document}